\documentclass[conference]{IEEEtran}
\IEEEoverridecommandlockouts
\usepackage{cite}
\usepackage{amsmath,amssymb,amsfonts}
\usepackage{algorithmic}
\usepackage{graphicx}
\usepackage{textcomp}
\usepackage{xcolor}
\def\BibTeX{{\rm B\kern-.05em{\sc i\kern-.025em b}\kern-.08em
    T\kern-.1667em\lower.7ex\hbox{E}\kern-.125emX}}
\begin{document}
\bstctlcite{IEEEexample:BSTcontrol}

\title{Bidirectional Power Packet Transmission\\Using an Inductive Energy Buffer
\thanks{This work was partially supported by JSPS KAKENHI Grant No.24K17262.}
}

\author{\IEEEauthorblockN{Yoshito Imanishi}
\IEEEauthorblockA{\textit{Graduate School of Engineering} \\
\textit{Kyoto University}\\
Kyoto, Japan
}
\and
\IEEEauthorblockN{Shiu Mochiyama}
\IEEEauthorblockA{\textit{Graduate School of Engineering} \\
\textit{Kyoto University}\\
Kyoto, Japan \\
ORCID: 0000-0001-7126-9187}
}

\maketitle

\begin{abstract}
The typical common-bus structure and its inherent bus stabilization requirement can become a bottleneck for flexibility in power supplies for battery-powered autonomous systems, limiting scalability and plug-and-play capability. Power packetization overcomes this by eliminating the common bus through time-division multiplexing of physically isolated power flows. However, conventional power packet routers face two physical-layer limitations: uncontrollable transmission direction and limited controllability of the amount of energy per packet. This paper proposes a fully controllable bidirectional power packet transmission to address these limitations. We introduce an inter-router circuit featuring a parallel inductor as a temporary energy buffer, enabling power transmission in any direction with full control over the transferred energy. We derive the switching algorithm for the routers to perform this operation and verify its feasibility through experiments using prototype hardware. 
\end{abstract}


\section{Introduction} \label{sec:intro}

With the expanding application of battery-powered autonomous systems, represented by electric vehicles and autonomous mobile robots, advanced energy management in isolated power systems has become an urgent issue. Conventional architectures for these systems are based on continuous power supply via power converters connected in parallel to a common bus\cite{Emadi.etal-2006,Safayatullah.etal-2022}. Ensuring the stability of such power systems even under abrupt load fluctuations requires to maintain a constant bus voltage. Various stabilization techniques have been proposed in the literature, including passive damping \cite{Cespedes.etal-2011}, active damping \cite{Rahimi.Emadi-2009} , and nonlinear control \cite{Emadi.etal-2006,Andres-Martinez.etal-2021}. However, they inevitably entail specific trade-offs, such as degraded system efficiency, dependency on load parameters, and immense complexity and extensive computational cost. 

This stabilization challenge is further complicated by the growing presence of bidirectional power flows, ranging from four-quadrant drives to vehicle-to-grid connections. While these are critical for pursuing ultimate high efficiency or prolonged operation time, the dynamic change in power flow direction introduces severe theoretical and practical difficulties. Specifically, when the power flow reverses, the dynamic characteristics shift between positive damping and negative impedance, creating an asymmetric stability margin. Furthermore, as the number of these bidirectional nodes increases, their arbitrary plug-in/plug-out operations unpredictably alter the global coupling dynamics. Consequently, the prerequisite of bus stabilization acts as a fundamental bottleneck that hinders the system's scalability and plug-and-play capability integrating diverse energy sources and loads\cite{Peng.etal-2026,Wang.etal-2025a}. %e.g., \cite{,} 

%To meet this requirement, on the contrary, conventional designs also requires a sufficient margin for the power density of the battery. However, practical autonomous systems impose conflicting demands: extended operation time (requiring high total energy) and superior mobility (requiring minimized weight and volume)\cite{Reveles-Miranda.etal-2024,Ankar.K.P.-2024}. Satisfying these demands under physical constraints inevitably sacrifices the power density margin. Conversely, restricting the dynamic characteristics of the load to prevent voltage instability significantly degrades the mobility of the system.

To overcome the fundamental bottleneck arising from bus stabilization, this paper considers an energy management approach based on power packetization \cite{Takuno.etal-2010,mochiyamaPowerPacketDispatching2021}. This method \textit{packetizes} power flows by discretizing them in the time domain and attaching information tags physically. The packetized units of power supply are individually routed to each load using time-division multiplexing according to the information tags, through a network composed of \textit{power packet routers}\cite{Takahashi.etal-2015,Yoshida.etal-2020}. This approach inherently eliminates the need for a common bus and thereby frees the system from stabilization requirements. 

In this paper, we develop physical layer technologies for realizing the transmission of packetized power between routers. The router proposed in previous studies \cite{Takahashi.etal-2015,Yoshida.etal-2020} suffers from two main issues arising from its circuit configuration (see Section~\ref{sec:packet} for detailed discussion). First, the direction of power transmission cannot be selected arbitrarily due to the restriction arising from routers' initial internal state. Second, the amount of energy that can be conveyed over a single power packet cannot be determined arbitrarily. The method proposed in this paper resolves these two issues.

The main contributions of this paper are as follows. First, we propose a novel circuit configuration between adjacent power packet routers (denoted hereafter by an inter-router circuit). Then, the corresponding switching algorithm for both routers is derived through analyses of circuit equations. We then confirm that the developed hardware and software realize power packet transmission that resolves the aforementioned issues. Lastly, we demonstrate the feasibility of the proposed method through experiments using a prototype implementation.

\section{Power Packetization and Routing} \label{sec:packet}

\begin{figure*}
    \centering
    \includegraphics[width=\linewidth]{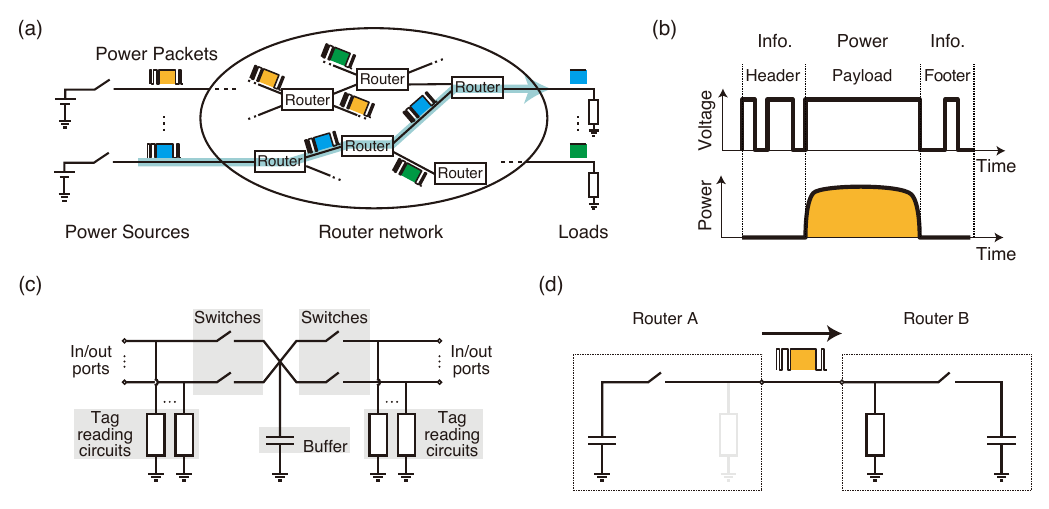}
    \caption{Concept and implementation of power packetization and routing\cite{Takuno.etal-2010}. (a) Generalized configuration of a power supply system based on power packetization. Power is packetized at sources, transferred through a network of routers, and arrives at loads. (b) Typical configuration of a power packet. The header and footer are information tags in voltage signals, accompanying the payload, the medium of pulsed power transmission. (c) Basic circuit configuration of the power packet router proposed in the previous studies\cite{Takahashi.etal-2015,Yoshida.etal-2020}. The number of ports and buffers can be varied arbitrarily by changing the switches' matrices. (d) Closed circuit formed when a power packet is delivered from a port of a router to a port of another router. }
    \label{fig:packet}
\end{figure*}

This section outlines the principles of power packetization and routing technologies based on previous studies. Furthermore, we clarify the limitations of the router circuits proposed in literature, thereby establishing the necessity of the bidirectional transmission method proposed in this paper. 

Fig.~\ref{fig:packet} illustrates the general configuration of a power system based on power packetization proposed in the literature\cite{Takuno.etal-2010,Takahashi.etal-2015,mochiyamaPowerPacketDispatching2021}. Fig.~\ref{fig:packet}~(a) presents the overview of the power system consisting of sources, loads, and a network of routers. Power from the sources is delivered to loads in a packetized form through the router network. 

Fig.~\ref{fig:packet}~(b) shows an example of a power packet configuration. A power packet is a unit of transmission that combines a DC power pulse, called a payload, with an information tag delivering signals using voltage waveforms. The information tag placed immediately before the payload, called the header, indicates the start of the packet and carries various information such as the source and destination of the power and control signals for the target load. Following the payload, the other tag, called the footer, indicates the end of the packet. 

Power supply to the load at the network edge is represented by a time series of the payloads. That is, the power flow is discretized in both time and quantity. This fundamentally differs from the conventional continuous power supply. Nevertheless, previous studies \cite{Mochiyama.Hikihara-2019a,Takahashi.etal-2016a} demonstrated that appropriately regulating the supply density of power packets can achieve an equivalent control to conventional PWM techniques. 

Power transmission from the source to the load is generally realized through a network of power packet routers. Fig.~\ref{fig:packet}~(c) depicts the router circuit proposed in \cite{Yoshida.etal-2020}. The router consists of a tag reading circuit, an temporary energy buffer, and a switching circuit. The tag reading circuit extracts the voltage envelop of the header of an incoming power packet and send the decoded information to the router's controller. Based on the read signal, the router executes information processing such as determining whether to accept the power packet and selecting the next transmission path. If the router decides to accept the power packet, the switch of the corresponding port is turned on and the payload power is stored in the temporary buffer. Then, by controlling the switch of the outgoing port, the router regenerates an information tag and a payload and forwards it to the next router. 

The previous study\cite{Yoshida.etal-2020} showed that adopting bidirectional switches such as back-to-back MOSFETs allows the router's ports to work as either an input or an output and thereby constitutes a bidirectional power packet router. However, there are still two limitations to a fully bidirectional transmission of power packets. 
First, the transmission direction cannot be determined arbitrarily. This limitation arises from the inter-router circuit configuration during a payload; power transmission relies on parallelizing capacitors of adjacent routers, as shown in Fig.~\ref{fig:packet}~(d). Since charge flows from higher to lower potential, the transmission direction is uniquely determined by the initial capacitor voltages at the onset of transmission.
The principle of charge redistribution between adjacent routers also poses the second limitation on the amount of payload power. During a payload, the current converges to zero when the voltages of both capacitors equalize. Therefore, it is fundamentally impossible to transmit all the energy stored in the sender-side capacitor as a single power packet.
To overcome these limitations, the subsequent section proposes a fully bidirectional power packet transmission method.

%Lastly, before moving to the proposed method, we briefly discuss the difference between the power routing based on physical packetization and other proposals. \textcolor{red}{Discuss the difference from: Solid-State Transformer\cite{She.etal-2013}, Multi-port DC Router\cite{Chen.etal-2020}, Energy Router}.

\section{Bidirectional Power Packet Transmission} \label{sec:bidir}

\subsection{Circuit Configuration}
\begin{figure}
    \centering
    \includegraphics[width=1\linewidth]{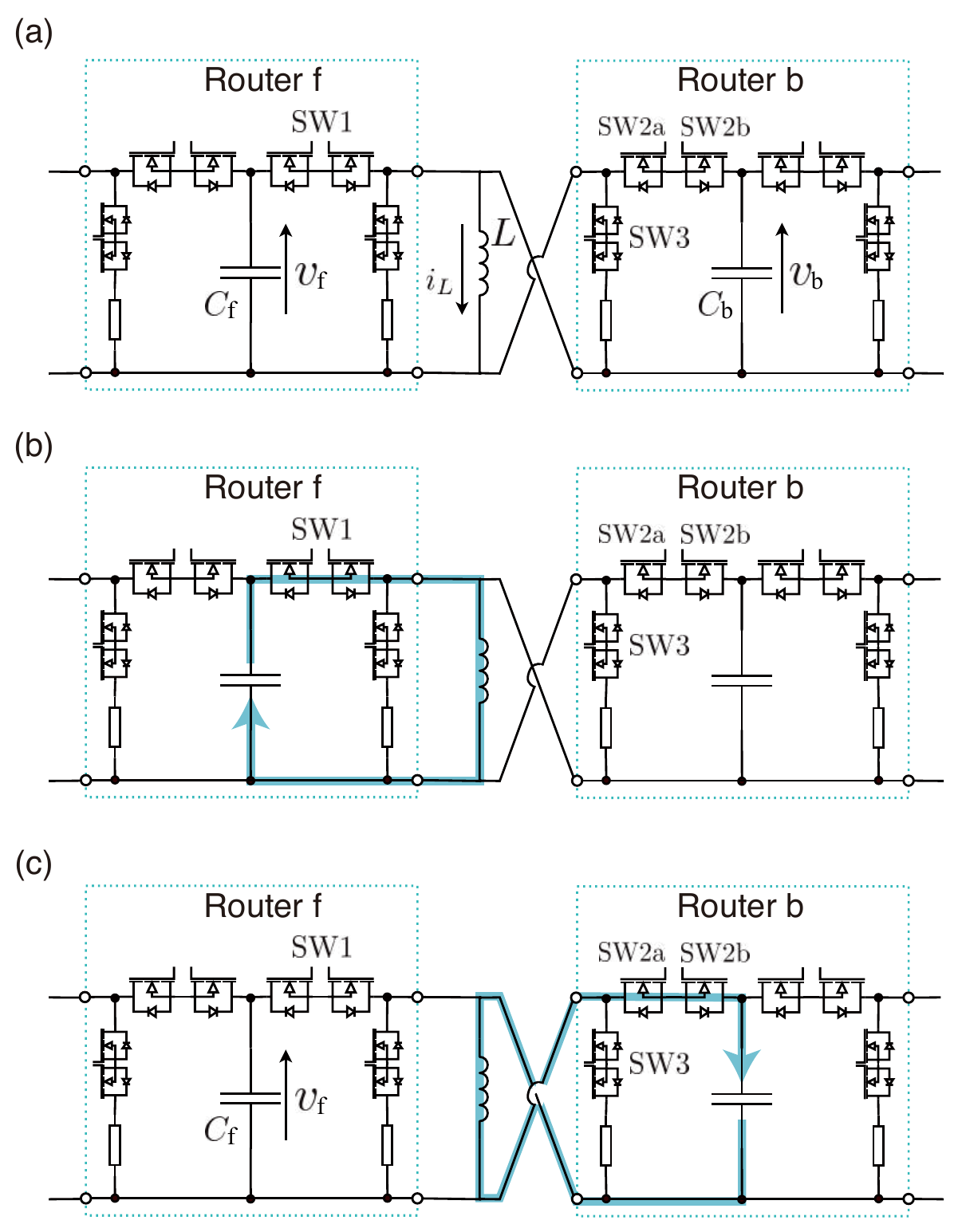}
    \caption{(a) Circuit configuration for the power packet transmission in the proposed method. The inductor is connected in parallel between the routers, while the routers themselves have the same circuit as in the previously proposed one. (b) Current path in the payload-f period. (c) Current path in the payload-b period. }
    \label{fig:circuit}
\end{figure}

The transmission between adjacent routers proposed in this paper is realized by the circuit shown in Fig.~\ref{fig:circuit}~(a). This circuit features an inductor placed in parallel between the ports of two adjacent routers. As detailed later, this inductor acts as temporary energy storage to enable bidirectional payload transmission. The routers themselves have the same circuit configuration with the previously proposed one (Fig.~\ref{fig:packet}~(c)). Based on the symmetry of the circuit, the following discussion assumes that the router on the left and right sides of the figure are designated as the sender and the receiver, respectively, without loss of generality.

Here, we focus solely on the transmission between two adjacent routers and do not consider the case where three or more routers form a closed circuit. This constraint is intended to ensure the scalability of the power system based on power packetization. Coordinating power packet transmission over multiple and spatially distributed routers requires strictly synchronized communication and control and thus necessitates a centralized controller. Generally, such a centralized architecture suffers from communication and computational bottleneck as the network size scales. Instead, we assume a decentralized architecture where each router operates autonomously while exchanging information only with its neighbors via an inforamtion tag. For this purpose, we define the power packet transmission as a localized process between two adjacent routers. 

%In this paper, we assume that the capacitance of each router is identical. That is, for the transmission circuit shown in Fig.~\ref{fig:circuit}~(a), we set $C_\mathrm{f}=C_\mathrm{b}=C$. It should be noted that, while this assumption is introduced to greatly simplify the discussion below, it is not essential for any of the subsequent discussion. A similar discussion is possible even when capacitance of routers are different. 

\subsection{Transmission Protocol}
We define the power packet transmission protocol utilizing the circuit introduced in the previous subsection.

\begin{figure}
    \centering
    \includegraphics[width=1\linewidth]{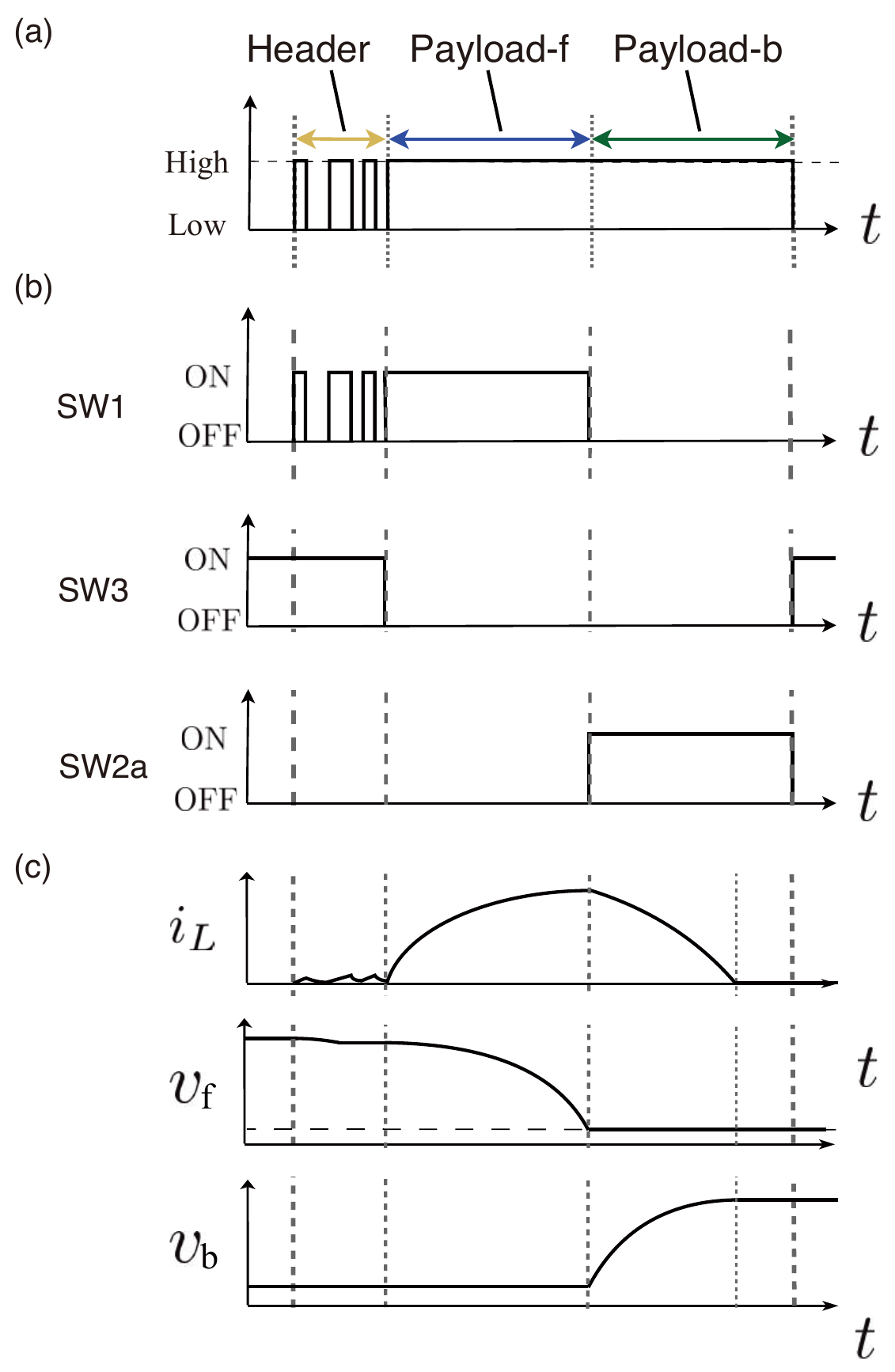}
    \caption{Protocol for the proposed power packet transmission. (a) Configuration of a power packet. The only difference from the previous studies is the division of the payload into two periods, denoted by payload-f and payload-b. (b) Switching chart for the routers' MOSFETs. The labels of the switches are identical to those in Fig.~\ref{fig:circuit}. (c) Representative waveforms of voltage and current when the switching of (b) is applied. The labels of voltage and current are identical to those in Fig.~\ref{fig:circuit}. }
    \label{fig:protocol}
\end{figure}

First, the structure of the power packet is defined as shown in Fig.~\ref{fig:protocol}~(a). The primary difference from the general configuration described in the previous section is that the payload is divided into two distinct periods, denoted by payload-f and payload-b. This is because the proposed power packet transmission comprises two stages: storing energy from the sender router into the inductor (payload-f) and discharging it from the inductor to the receiver router (payload-b). We denote the duration of the payload-f and payload-b by $T_\mathrm{f}$ and $T_\mathrm{b}$, respectively. Similar to the general configuration, the header can include various signals including the destination, but the specific setups are beyond the scope of this paper. The footer is omitted because the packet length is fixed at the prior design as presented below. 

Fig.~\ref{fig:protocol}~(b) illustrates the timing chart of switches' states and the schematics of resulting voltage and current waveforms. First, during the header period, the sender router transmits logical values as a voltage waveform by toggling SW$_1$ on and off. The receiver router reads this via its reading circuit by turning on SW$_3$. The header transmission process is completely same as defined in the previous proposals\cite{Takahashi.etal-2015,Yoshida.etal-2020}. Then, in the subsequent payload-f period, the sender router turns on only SW$_1$ to store the energy of the capacitor $C_\mathrm{f}$ into the inductor. The current path in this period is depicted in Fig.~\ref{fig:circuit} ~(b). Finally, in the payload-b period, the sender router turns off SW$_1$, and the receiver router turns on SW$_\mathrm{2a}$, releasing the stored energy from the inductor to the capacitor $C_\mathrm{b}$. By turning on SW$_\mathrm{2a}$ only, the diode parallel to the SW$_\mathrm{2b}$ blocks reverse current once the inductor current reaches zero, preventing power flowing from the capacitor back to the inductor. The current path in this period is depicted in Fig.~\ref{fig:circuit}~(c). 
%Thus, the maximum duration of the payload-b period, $T_b$, defined as the time until the inductor current drops to zero, can be uniquely determined from the circuit equations.

In summary, by employing the inductor as a temporary energy buffer, the proposed method isolates the sender and receiver routers in time and forms two separate LC circuits. The decoupling of the routers enables power transmission in an arbitrary direction regardless of the initial voltage levels of the capacitors. Furthermore, by ensuring the inductor current reaches zero at the end of the payload-b transmission, no energy remains in the inductor, thereby isolating energy on a per-packet basis. This is essential to maintain the core principle of the power packetization, the distinct power flow separation, by avoiding the different power flows from getting mixed.

Next, we describe the configuration for $T_\mathrm{f}$ and $T_\mathrm{b}$. First, $T_\mathrm{f}$ is determined by the target amount of energy to be transferred. For this, we employ the transmission ratio $\mu \in (0,1]$, defined as the ratio of the total energy extracted from the sender-side capacitor over the payload-f duration to its initially stored energy. Then, $T_\mathrm{b}$ is set such that the energy stored in the inductor at the end of payload-b is zero. In general, for a given set of circuit parameters, $L,C$, the exact time when the inductor current reaches zero (denoted as $\tau$) depends on operating conditions such as $\mu$ and the initial voltage of the router capacitors. Here, we define $T_\mathrm{b}$ as the maximum possible value of $\tau$ across all allowable operating conditions. Since the parallel diode in SW$_\mathrm{2b}$ blocks the reverse current, zero current is maintained for any duration $T_\mathrm{b} \ge \tau$. Setting $T_\mathrm{b}$ to this universal maximum provides a robust safety margin against uncertainty in operating conditions.

Based on the above discussion, specific values for $T_\mathrm{f}$ and $T_\mathrm{b}$ are derived by solving initial value problems of the circuit equations as follows (see Appendix for their derivation):
\begin{align}
 T_\mathrm{f} &= \frac{1}{\omega_\mathrm{f}} \arcsin \sqrt{\mu}, \label{eq:tf} \\
 T_\mathrm{b} &= \frac{\pi}{2} \frac{1}{\omega_\mathrm{b}}, \label{eq:tb} \\
 \omega_\mathrm{f} &= \sqrt{\frac{1}{C_\mathrm{f}L} - \left(\frac{r_\mathrm{f}}{2L}\right)^2}, \\
 \omega_\mathrm{b} &= \sqrt{\frac{1}{C_\mathrm{b}L} - \left(\frac{r_\mathrm{b}}{2L}\right)^2},
\end{align}
where $r_\mathrm{f}$ and $r_\mathrm{b}$ represent the sum of the parasitic resistance on the sender and receiver sides, respectively, including the inductor's DC resistance and the on-resistance of the MOSFET(s) and the diode.

\section{Experiments}

In this section, we present the proposed power packet transmission using a prototype circuit and an example setup of parameters. To focus on the essential part of the proposed method, we omit the transmission of information tags, which is done in the same way as the previous studies. 

\subsection{Setup}

\begin{figure}
    \centering
    \includegraphics[width=1\linewidth]{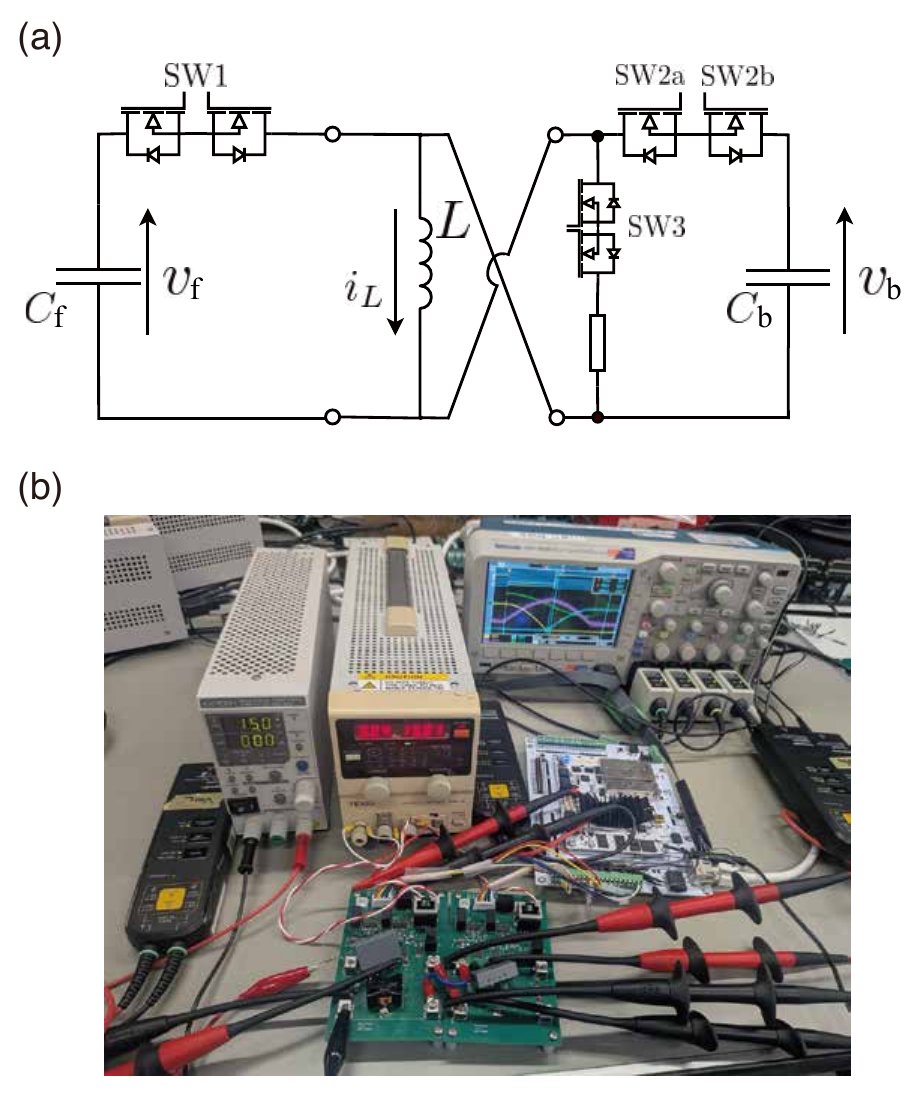}
    \caption{Experimental setup. (a) Circuit configuration for the experiment. This circuit captures an essential part of Fig.~\ref{fig:circuit}. (b) Photograph of the implemented setup.}
    \label{fig:experiment}
\end{figure}

We fabricated the circuit corresponding to a pair of routers as shown in Fig.~\ref{fig:experiment}~(a). Considering the symmetry of the circuit and its operation discussed in Section~\ref{sec:bidir}, we focus on the transmission from the left-hand side router to the right-hand side router only, without loss of generality. 
The bidirectional switches were realized by back-to-back MOSFETs. The routers' energy buffers were both implemented with film capacitors of nominal capacitance $10\,\mu\mathrm{F}$. We identified their actual capacitance using an impedance analyzer: $C_\mathrm{f}=10.01\,\mu\mathrm{F}$ and $C_\mathrm{b}= 10.04\,\mu\mathrm{F}$. We adopted the film capacitors for their better DC-bias characteristics than ceramic capacitors. The inter-router buffer was implemented with a power inductor of nominal inductance $47\,\mu\mathrm{H}$, whose actual inductance was $L=49.89\,\mu\mathrm{H}$. The inductor was selected with sufficient saturation current ratings. 

Substituting the actual values of $C_\mathrm{f},C_\mathrm{b},L$ to Eq.~(\ref{eq:tf})~and~(\ref{eq:tb}), the durations of payload-f and payload-b were calculated to be $T_\mathrm{f}=35.10\,\mu\mathrm{s}, T_\mathrm{b}=35.16\,\mu\mathrm{s}$, where we assumed $\mu=1$, namely the case where all the energy stored in the sender capacitor is transferred to inductor during payload-f. In the calculation, we used the estimated values for parasitic resistance of the circuit during payload-f and payload-b, $r_\mathrm{f}$ and $r_\mathrm{b}$. The estimates were determined based on the datasheet values of the circuit components used in the experiment; the specific values were determined as $r_\mathrm{f}=25.22\,\mathrm{m}\Omega$ and $r_\mathrm{b}=28.21\,\mathrm{m}\Omega$. 

To realize the sub-$\mu\mathrm{s}$ timing constraints of the circuit switching, we employed the user-programmable logic of the DSP controller (B-Board; Imperix). 
The overall implementation of the circuit, controller, and other apparatuses is shown in Fig.~\ref{fig:experiment}~(b). 

\subsection{Results and Discussion}

\begin{figure}
    \centering
    \includegraphics[width=1\linewidth]{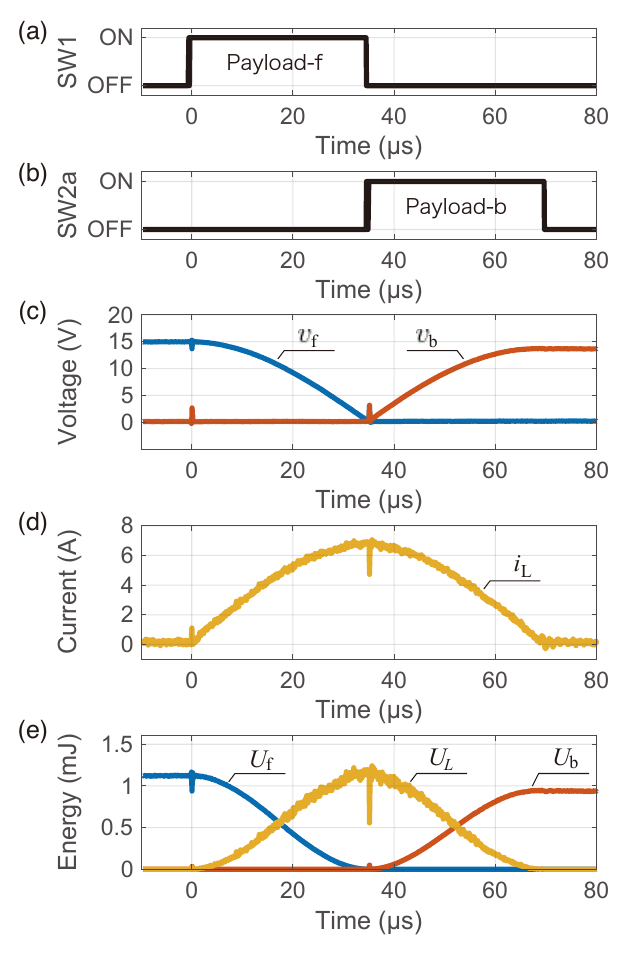}
    \caption{Results of the experiment. (a) Switching state of SW$_1$. (b) Switching state of SW$_\mathrm{2a}$. (c) Voltage waveforms of the capacitors. (d) Current waveform of the inductor. (e) Energy waveforms for the capacitors and inductor. The energy values are estimations calculated by using the capacitance and inductance values and measured voltage and current. }
    \label{fig:result}
\end{figure}

Using the obtained payload durations, we conducted an experiment of transmission of one payload. We measured voltage across the capacitors, $v_\mathrm{f}, v_\mathrm{b}$, and inductor current, $i_L$. The energy stored in the sender and receiver capacitors are denoted by $U_\mathrm{f}$ and $U_\mathrm{b}$, respectively. We present the estimated values for $U_\mathrm{f}$ and $U_\mathrm{b}$ based on 
\begin{equation}
    U_\mathrm{x} = \frac{1}{2}C_\mathrm{x} {v_\mathrm{x}}^2 \quad (\mathrm{x=f,b}).
\end{equation}
Similarly, we denote the energy stored in the inductor by $U_L$, and its estimate is calculated by
\begin{equation}
    U_L = \frac{1}{2}L {i_L}^2.
\end{equation}

First, we observe the switching operations in payload-f and payload-b periods. Fig.~\ref{fig:result}~(a)~and~(b) present the transitions of the switching states of the switches, SW$_1$ and SW$_\mathrm{2a}$, respectively. Fig.~\ref{fig:result}~(c)~and~(d) present the voltage and current waveforms, respectively. While SW$_1$ were on (during payload-f), $v_\mathrm{f}$ gradually decreased. At the same time, $i_L$ increased. These results indicate that energy transferred from $C_\mathrm{f}$ to $L$. Then, at the timing when $v_\mathrm{f}$ reached zero, SW$_1$ turned off and SW$_\mathrm{2a}$ turned on, entering payload-b. The transitions of the switching states triggered the circuit's operation mode; current through $L$ started decreasing while $v_\mathrm{b}$ increasing. Lastly, when SW$_\mathrm{2a}$ turned off, current through $L$ reached zero, which means that energy temporarily stored in $L$ transferred to $C_\mathrm{b}$. 
%Here, we can see an oscillation occurred at the end of the paylaod-b period; we discuss this point later in this section. 

Next, we observe the transition of energy in payload-f and payload-b periods. Fig.~\ref{fig:result}~(e) presents the variation of energy in the capacitors and inductor with respect to time. In payload-f period, the sender's energy $U_\mathrm{f}$ gradually transferred to the inductor. Then, in payload-b period, the buffered energy $U_L$ transferred to the receiver's capacitor, which completed the energy transfer from the sender router to the receiver router. 

%In the energy transition, we observed that the energy in $C_\mathrm{f}$ becomes small but not exactly zero at the intersection of the payload-f and payload-b. In fact, Fig.~\ref{fig:result}~(b) presents that the switching between the payload-f and payload-b periods occurred before voltage across $C_\mathrm{f}$ reached zero. This was due to the following two reasons. First, since we determined $T_\mathrm{f}$ by calculating with the nominal value of $L,C$, the variation in their actual values can cause a slight difference in $T_\mathrm{f}$. Second, the limited sampling rate of the controller constrained the timing accuracy of the circuit switching. The parameter variation can be addressed by introducing voltage measurement across the router capacitor and adjust the switching timing based on the measurement. Then, the restricted timing accuracy of the control can be resolved simply by adopting a DSP with higher sampling rate. Applying these solutions to improve the performance of the proposed method would be a future work. 

We observed the oscillation in the inductor current after the end of the payload-b. This should be addressed because the lasting oscillation interferes the transmission of a next power packet. We consider that the oscillation is due to the resonance of the inductance and the parasitic capacitance of the switches. An effective way to suppress the oscillation would be adding a snubber circuit to the switches\cite{Mohan-etal-2003}, which forms an important future work. 

%The power packet transmission inevitably produces energy loss. In Fig.~\ref{fig:}~(a), the difference between the initial value of the energy in $C_\mathrm{f}$ and the peak value of the energy in $L$ represents the loss during the payload-f period. Similarly, the difference between the peak value of the inductor energy and the steady state value of the energy in $C_\mathrm{b}$ represents the loss during the payload-b period. The specific value of the total loss during one packet transmission was \textcolor{red}{$??\,\mu\mathrm{J}$}. This was in fact larger than estimated from a prior circuit analysis based on datasheet values of parasitic resistance. 

\section{Conclusions}
In this paper, we developed the physical layer technology for bidirectional power packet transmission between adjacent routers. 
The proposal consists of both hardware and software improvement of the power packet routing in the literature, removing their restriction in the controllability for direction and amount of power transmission. 
We demonstrate the basic operation of the developed technology through experiments using a prototype implementation. 
Based on the experimental results, we also discussed issues in the current implementation, which we address in the future research. 

\section*{Appendix}

We assume that the parasitic resistance of the transmission lines and switches is sufficiently small, and the circuit satisfies the underdamped condition:
\begin{equation}
    \alpha_\mathrm{f} < \beta_\mathrm{f}, \quad \alpha_\mathrm{b} < \beta_\mathrm{b},
\end{equation}
where
\begin{align}
\alpha_\mathrm{f} &= \frac{r_\mathrm{f}}{2L},\quad
\alpha_\mathrm{b}  = \frac{r_\mathrm{b}}{2L},\quad
\beta_\mathrm{f} = \sqrt{\frac{1}{C_\mathrm{f}L}},\quad
\beta_\mathrm{b} = \sqrt{\frac{1}{C_\mathrm{b}L}}.
\end{align}

First, Eq.~(1) is derived. Let $t=0$ be the start time of payload-f, with initial conditions $v_\mathrm{f}(0)=V_\mathrm{f}$ and $i_L(0)=0$. Assuming the parasitic resistance is negligible, solving the circuit equation yields 
\begin{equation}
    v_\mathrm{f}(t)= V_\mathrm{f}\cos\omega_\mathrm{f} t .
\end{equation}
Substituting this into the relationship specifying the target energy transmission ratio $\mu$, given by
\begin{equation}
Cv_\mathrm{f}(T_\mathrm{f})^2 = (1-\mu) C {v_\mathrm{f}(0)}^2 ,    
\end{equation}
yields 
\begin{equation}
    \sin(\omega_\mathrm{f}T_\mathrm{f})=\sqrt{\mu}.
\end{equation}
Solving this for $T_\mathrm{f}$ gives Eq.~(1).

Next, (2) is derived. Resetting $t=0$ to the start of payload-b, the initial conditions are set to $v_\mathrm{b}(0)=V_\mathrm{b}$ and $i_L(0)=I_\mathrm{b}$. To account for the body diode voltage, let $\tilde{v}_\mathrm{b}=v_\mathrm{b}+V_\mathrm{F}$. Solving the circuit equation yields
\begin{align}
i_L(t) &=
e^{-\alpha_\mathrm{b} t} \left\{
-\frac{\tilde{V}_\mathrm{b}}{L\omega_\mathrm{b}}\sin\omega_\mathrm{b} t
+\frac{\beta_\mathrm{b} I_\mathrm{b}}{\omega_\mathrm{b}}\cos(\omega_\mathrm{b} t+\phi_\mathrm{b})
\right\},
\label{eq:a_b}
\end{align}
where $\cos\phi_\mathrm{b}=\omega_\mathrm{b}/\beta$ ($0<\phi_\mathrm{b}<\pi/2$). Due to the body diode of SW$_\mathrm{2b}$, the inductor current is blocked when it first reaches zero, which we denote by $\tau$. Substituting $i_L(\tau)=0$ into Eq.~(\ref{eq:a_b}), $\tau$ is obtained as
\begin{equation}
\tau = \frac{1}{\omega_\mathrm{b}}\arctan\left\{
\frac{L\omega_\mathrm{b} I_\mathrm{b}}{\tilde{V}_\mathrm{b}+\alpha_\mathrm{b}L I_\mathrm{b}}
\right\}.
\label{eq:tau}
\end{equation}
The duration $T_\mathrm{b}$ must be designed such that $\tau\le T_\mathrm{b}$ holds under any initial conditions. Since $I_\mathrm{b}>0$ and $\tilde{V}_\mathrm{b}>0$, the argument of the arctangent in Eq.~(\ref{eq:tau}) is strictly positive. Considering the range of the arctangent function, $(0,\pi/2)$, it follows that
\begin{equation}
0 < \omega_\mathrm{b}\tau < \frac{\pi}{2},
\end{equation}
which yields Eq.~(2). 

%\section*{Acknowledgment}

\bibliographystyle{IEEEtran}
\bibliography{references.bib}

\end{document}